# Enhancing the limit of uniaxial magnetic anisotropy induced by ion beam erosion.

Anup Kumar Bera, Arun Singh Dev and Dileep Kumar [a)]

*UGC-DAE Consortium for Scientific Research, Khandwa Road, Indore-452001, India*

[a)] Corresponding author: dkumar@csr.res.in

The artificial tailoring of magnetic anisotropy by manipulation of interfacial morphology and film structure are of fundamental interest from application point of view in spintronic and magnetic memory devices. This letter reports an approach of engineering and enhancing the strength of oblique incidence ion beam erosion (IBE) induced in-plane uniaxial magnetic anisotropy (UMA) by simultaneous modification of film morphology as well as film texture. To meet this objective, Cobalt film and Si substrate have been taken as a model system. Unlike conventional post growth IBE of film, we direct our effort to the sequential deposition and subsequent IBE of the film. Detailed in-situ investigation insights that the film grows in highly biaxially textured polycrystalline state with formation of nanometric surface ripples. The film also exhibits pronounced UMA with easy axis oriented parallel to the surface ripple direction. Remarkably, the induced UMA is about one order of magnitude larger than the reported similar kind of earlier studies. The possibility of imposing in-plane crystallographic texture giving rise to magneto-crystalline anisotropy, along with long-range dipolar interaction throughout ripple crests enhances the strength of the UMA. The present findings can be further extended to systems characterized by different crystallographic structure and magnetic properties and show the general applicability of the present method.

## INTRODUCTION

Magnetism in low dimensional systems such as magnetic nano-structures, ultrathin film and multilayers are of great interest from both fundamental and applied point of view and have become an active field of research. In particular, uniaxial magnetic anisotropy (UMA) and magnetization reversal are the key properties which finds its application in high-density magnetic memories,[1] magnetic sensors,[2] spintronic and wireless communication devices.[3] Therefore, researchers have been developing several strategies such as oblique angle deposition,[4] magnetic field and stress annealing,[5] engineering of interface[6] etc. for flexible and fine tailoring of UMA. In this respect ion beam erosion (IBE) has been demonstrated as a versatile and handy tool, in particular way to induce UMA by engineering surface and interface morphology through self-assembled formation of nanometric pattern on the surface.[7–12] Ferromagnetic films deposited on prepatterned nano structured substrate[7–9] (bottom up approach) or post growth IBE of the film surface[10–12] (top down approach) deposited on planner substrate is found to imprint an UMA. However, these studies suffer from its own limitations. For example in top down approach, ion beam induced modifications happens only on the top surface due to the limited penetration depth of ions.[12] Moreover, initial film thickness must be high enough for homogeneous pattern formation and also for maintaining film continuity.[13] On the other hand, in bottom up approach the strength of UMA fall off after a critical film thickness due to merging of ripples crests with their nearest neighbors.[7,14,15] Furthermore, due to presence of randomly oriented grains in polycrystalline film it is always a challenging task to induce magneto crystalline anisotropy (MCA) by modification of film texture through preferred grain alignment. However, earlier it has been observed that due to minimization of magneto-elastic energy and coupling of MCA with shape anisotropy, c-axis gets oriented either perpendicular to ripple or parallel to the long axis of column and nanowire.[8,16] Therefore, one promising approach would be inducing crystallographic texture aligned parallel to the ripple direction for intensifying UMA. Despite being aware of the fact that the interaction between ion beam and thin film can alter the structure or crystallinity of material,[17] no studies in the literature have taken this contribution into account to enhance UMA in thin films. This may be owing to the ex-situ characterization where surface oxidation and contamination destroys film surface structure. Therefore, in order to extract genuine and unambiguous information about surface structure, film morphology and its correlation with magnetic anisotropy use of surface sensitive techniques and more importantly in-situ characterization is highly required.

Within this context, we have followed a sequential deposition and IBE procedure for tuning of both the crystallographic texture and morphology simultaneously to achieve strong UMA. In-situ reflection high energy electron diffraction (RHEED) and X-ray diffuse scattering (XDS) provided information about film structure and morphology, while magneto optical Kerr effect (MOKE) gave information about magnetic anisotropy, thus making it possible to correlate the film structure and morphology with that of UMA in the film. Furthermore, unlike the conventional "top down" approach, the present method permits textured growth throughout the whole layer of the film. Thus, the film exhibits strong UMA originating from combined effect of MCA and shape anisotropy.

The sample preparation and characterization are done inside an ultra-high vacuum (UHV) chamber having a base pressure ≈5× $10^{-10}$ mbar or better. It is equipped with facilities for growth of thin film using electron bean evaporation technique and in-situ characterization using MOKE, RHEED, X-ray reflectivity (XRR) and XDS measurements. In the present study a Si (001) wafer covered with a native oxide layer is used as substrate. Co film of thickness ≈20nm is deposited in 20steps on this substrate with a deposition rate of ≈0.2 Å/s as readout by a water-cooled calibrated quartz crystal monitor. Each step of sequential deposition-erosion process consists of 1nm thick Co film deposition followed by 2 minutes of film surface erosion using 2keV Ar+ ion beam at an angle ~50° from the surface normal. The ion flux on the sample was approximately $10^{12}$ ions/cm$^2$ as separately measured by a faraday cup. MOKE, RHEED, XRR and XDS were performed at few intermediate steps as well as after complete deposition. In order to draw intercorrelation among structural, morphological and magnetic anisotropy in conjugation with the direction of IBE, all the measurements are performed by rotating the sample with respect to the IBE direction.

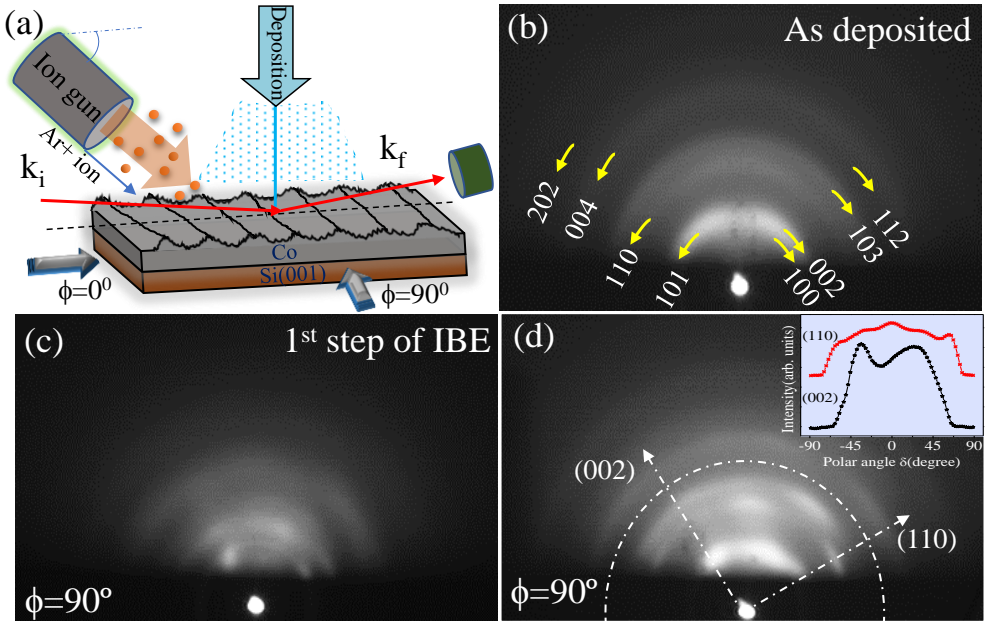

**Fig 1**: (a) Schematic of experimental geometry. (b) In-situ RHEED images taken at 90° with respect to the IBE direction after 1nm film deposition. RHEED images taken (c) along and (d) across to the IBE direction after 20$^{th}$ step of deposition. (Inset) The corresponding polar intensity profile extracted from the first ring of RHEED image along $\phi$=90°. The dotted line with arrow head in the RHEED pattern indicates the surface normal of planes.

A schematic of experimental geometries with all relevant equipment and directions are presented in Fig 1(a). Figure 1(b) represents the RHEED diffraction pattern of the film taken after deposition of 1nm thick Co film. The diffraction pattern consists of symmetric, concentric and continuous Debye rings with uniform intensity distribution. This is a signature of random nucleation of crystallites which confirms polycrystalline nature of the film. The rings corresponding to different planes are identified from their position and marked in the Fig 1(b). It may be noted that the first ring from the center is quite broad compared to the other rings due to convolution of (101), (002) and (102) ring [18]. The corresponding RHEED image after 2min of IBE is given in the Fig 1(c). A drastic change in the ring pattern can clearly be seen in the image. It exhibits asymmetric and nonuniform intensity distribution with presence of broken arcs along the Debye rings. It confirms the existence of preferred orientation (texturing) of the different crystallites. The similar process is further followed for 20 steps which includes sequential film deposition and successive IBE. The RHEED image taken after whole process is shown in Fig 1(d). It is clear from the figure that the film is displaying pronounced texturing even after 20 steps of process. In order to give insight on the substantial arrangement of the distribution of ring intensity and texture axis, the radially integrated intensity distributions taken at the position of (002) ring is plotted with polar angle δ in the inset of Fig. 1(d) where δ is defined as the angle measured from the vertical axis perpendicular to the substrate (δ=0º is at top of the ring). This selection is driven by the fact that <002> direction is easy axis of MCA. We found that the [110] and [002] texture axis is tilted from the surface normal by ~60º and ~30º respectively which matches with the standard angle between two direction. The origin of the IBE induced texture formation is due to the difference in the degree of the ion channeling effect or ion-induced anisotropic radiation damage in the crystal plane[19] and epitaxial overlayer growth favoring the nucleation on grains with low sputter yield orientation. Thus, with the help of the present method crystallographic texture can be induced throughout the whole layer of film and the texture axis can be tilted away from normal of the film surface to the ion flux direction.[20]

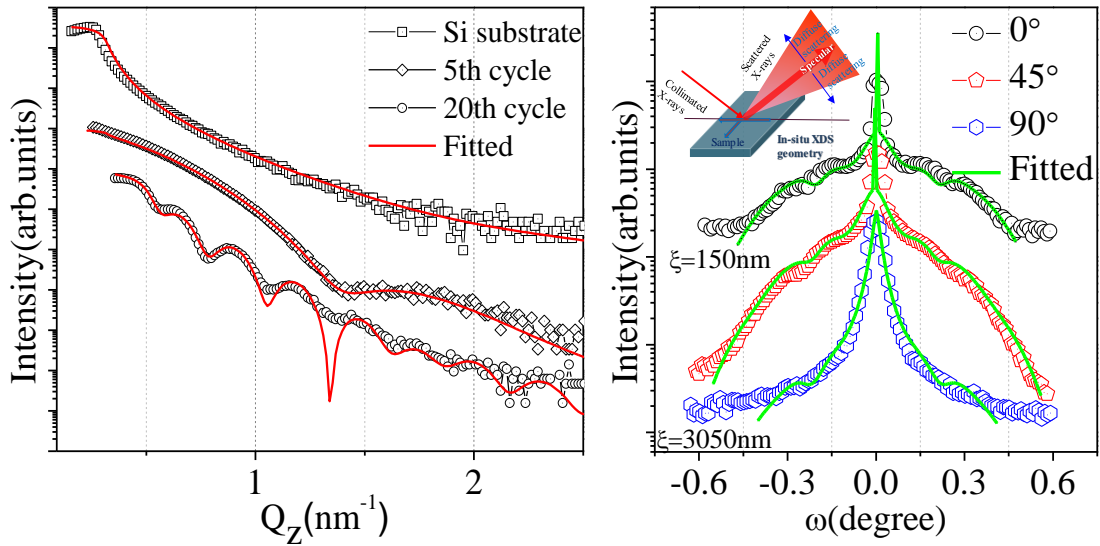

**Fig 2:** (a) In-situ XRR plot of Si substrate and Co film at the end of 5th and 20th cycle of deposition and IBE. The hollow lines represent the experimental data and the red lines represent the best fit to experimental data using Parratt formalism. (b) Plots of in-situ XDS measurement as a function of the rocking angle ω (=θi−θsc) along different azimuthal angle from IBE direction. The spectra have been shifted vertically for clarity purposes. Inset shows the schematic of XDS geometry.

In-situ X-ray scattering techniques, XRR and XDS have been used to study both the specular and diffuse scattering to get the in-plane and out plan surface morphology of the film. Figure 2(a) gives the XRR pattern of the bare Si substrate and after 5th and the 20th step of deposition-erosion process. The presence of periodic multiple peaks (Kiessig fringes) indicates good structural order and morphological uniformity within the film. XRR patterns have been fitted using Parratt's formalism.[21] The extracted root mean square (rms, $\sigma$) roughness and film thickness (d) are as follows: $\sigma$ =0.5 nm for substrate, d= 4.6nm and $\sigma$ = 0.6nm for film after 5th cycle, d=18nm, $\sigma$=0.9nm for film after 20th cycle.

To extract lateral correlations of the surface morphology, XDS measurements are performed along different azimuthal directions ($\phi$=0°,45°,90°, 0° is along the IBE direction) by keeping the scattering angle $\theta_{sc}$ fixed at 0.62° and varying the angle of incidence $\theta_i$ from 0° to 2$\theta_{sc}$.[11] As shown in Fig 2(b), along $\phi$=90°, XDS curve exhibits distribution of scattered intensity only near-specular region (at $\omega \approx$ 0). In contrast, small subsidiary peaks or maximas are present along $\phi$=0° on both side of specular reflection. The intensity oscillations correspond to a characteristics length scale of periodic surface morphology. To get quantitative information of surface morphology, XDS curves are fitted using TRDS_sl program in the framework of distorted wave born approximation formalism,[11] where the distribution of height and its correlation in the *x,y* plane is expressed by the height-height correlation function[22–24] $C(x,y) = \sigma^2 \exp(-|R|/\xi)^{2h}$, where $R = (x^2 + y^2)^{1/2}$; where, $\sigma$ is rms roughness, $\xi$ describes length scale of correlated lateral ordering on the surface and exponent *h* is jaggedness parameter. The fitted values of the parameters $\xi$ and h are found to be 150nm and 0.25 for $\phi$=0°, 3050nm and 0.42 for $\phi$=90°, respectively. Since $\xi$ corresponds to the length scale at which a point on the surface follows the memory of its initial value, the obtained values suggest that short and long axis of the anisotropic surface morphology is oriented along $\phi$=0° and $\phi$=90° respectively. It also confirms formation of periodic ripple like pattern with ridges running perpendicular to the IBE direction. In order to provide visual evidence of the formation of ordered rippled morphology, a separate film has been deposited in identical condition. The corresponding AFM image of the film surface is provided in the supplementary material. It exhibits periodic ripples with groves running perpendicular to the projection of IBE direction on the film surface and confirms formation of unidirectional correlated rippled morphology in the film that also complements the in-situ XDS measurement.

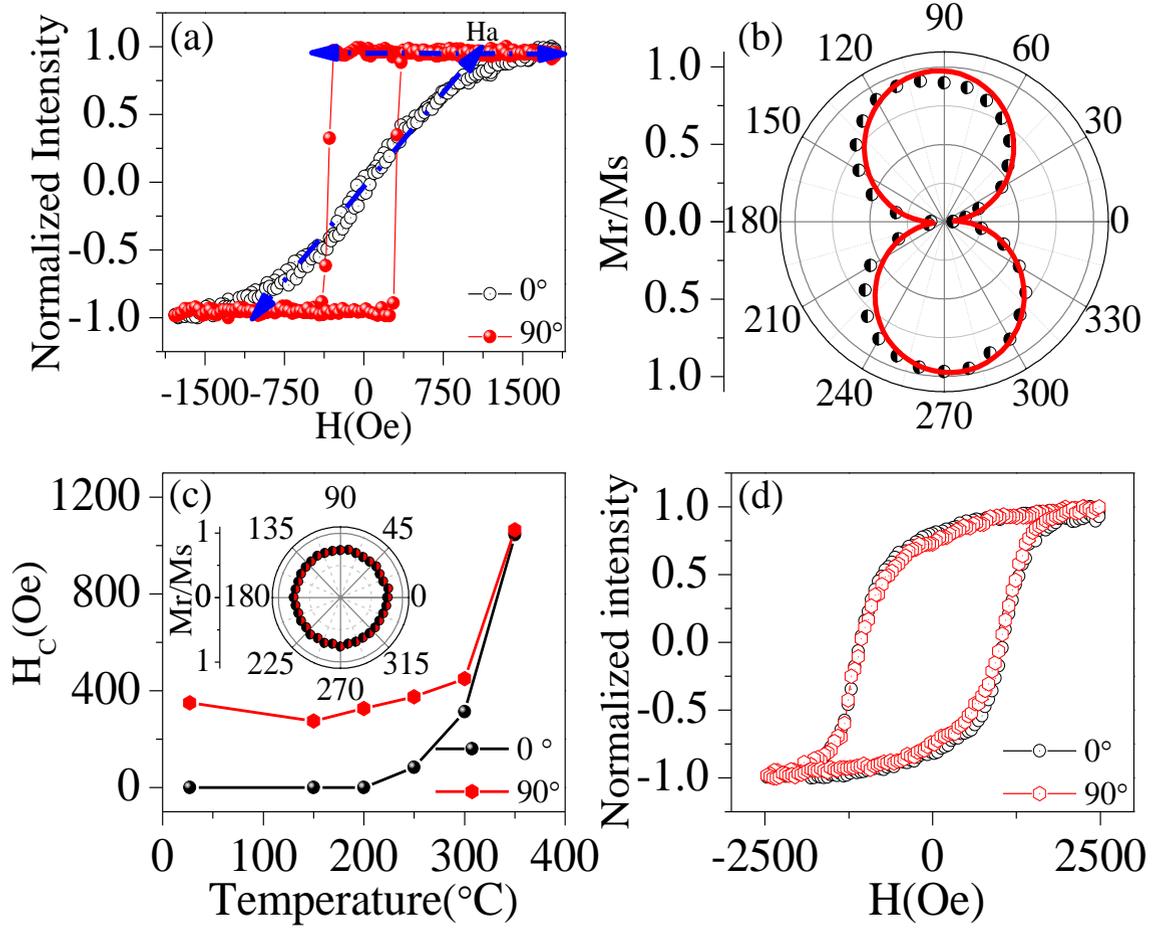

**Fig 3**: (a) In-situ MOKE hysteresis loops along ɸ=0° and ɸ=90° direction. (b) A polar plot of the normalized magnetization. (c) Variation of coercivity with annealing temperature. The inset shows the corresponding polar plot of normalized remanence at 350°C. (d) In-situ MOKE hysteresis loop along ɸ=0° and ɸ=90° direction taken after annealing at 350°C.

The magnetic characterization of the sample was carried out in-situ by MOKE measurements in longitudinal geometry and analysis was focused on determination of strength as well as orientation of the in-plane anisotropy axis. Figure 3(a) displays the representative MOKE hysteresis loops measured with external field H applied along ɸ=0° and ɸ=90° directions. The hysteresis curve along IBE direction displays no opening in the middle of the loop suggesting magnetization switching in this direction is reversible and proceeds via coherent rotation in the applied field direction. However, along ɸ=90° direction, hysteresis curve exhibits high squareness with coercive field $(Hc) \approx 400$ $Oe$. The full angular (0 to 360°) dependence of remanence magnetization $(Mr/Ms)$ is plotted in Fig. 3(b). It has been modeled with the following cosine like function: $\frac{M_r}{M_s} = a|\cos(\phi - \phi_0)| + b$ to extract exact orientation and degree of anisotropy. Here, $a$ define the degree of anisotropy, b arises due to isotropic contribution, $\phi_0$ is the angular offset between the easy axis and the zero setting of the rotational sample stage. From fitting the obtained value of $a$ is 94% and $\phi_0$ is 3º. Thus, a strong UMA is present with easy axis of magnetization oriented almost 90º to the IBE direction. In order to estimate the strength of UMA, the experimental value of UMA energy, $K_U$, has been calculated using the relation $K_U = M_s H_a/2$. Where, Ms is the bulk saturation value (1422 emu/cm$^3$) of Co. $H_a$ is the anisotropy field, which has been determined by the extrapolation of the linear magnetization regime obtained along the hard axis direction at the value of the saturation magnetization.[9] $Ha$ =1100 Oe is also marked by the blue line in Fig 3(a). The value of $Ku \approx 7.82 \times 10^5$ erg/cm$^3$ is obtained, which is one order of magnitude greater than the similar previous studies.[7–10,14] For instance, Sarathlal[8] et al. have obtained $Ha \approx 120$ Oe of 18nm thick Co film, Bukharia[25] et al. have obtained $Ha \approx 100$-150 Oe for Co with thickness ranging from 5-60nm, Liedke[7] et al. have obtained anisotropy field in the range 100-250 Oe for 2-40nm Co, Fe and Py film deposited over rippled substrate. J. Berendt[26] have obtained $Ha \approx 10$ Oe of 20nm thick permalloy film deposited on holographic grating. Zhang[12] et al. have achieved $Ha \approx 75$ Oe only by IBE of Fe film. Compared to these studies, almost one order magnitude higher $Ha \approx 1100$ Oe has been achieved. Till now in this field except the present study, Miguel[27] et al. have achieved the highest value of $Ha \approx 800$Oe by direct modifying the surface morphology of Co thin film having large ripple amplitude upto 20 nm as the leading variable. All this films, the film surface is modified by single erosion step. Therefore, only top surface is expected to be modified. No evidence of crystallographic texture could be due to the top surface contamination in ex-situ studies. Thus, it is evident that sequential film deposition and IBE of film imprints a strong UMA.

It is noteworthy that the magnetostatic contribution to the total strength of induced UMA for a magnetic film with anisotropic surface roughness is evaluated from morphological parameters using Schlömann's formula[28] $Ku(shape) = 2\pi M_S^2 \frac{\pi \sigma^2}{\lambda d} = 2\pi M_S^2 (Na - Nc)$ and a good agreement between theoretical and experimental values are obtained in several previous studies.[7,9,14] Here, Na and Nc are the demagnetization factor along short and long axis. $Ku = 1.4 \times 10^4$ erg/cm$^3$ has been obtained by putting the simulated values extracted from XDS measurement. This significant difference between the experimental value obtained from hysteresis loop and theoretical value calculated from Schlömann's formula indicates that the induced UMA not solely originating from shape anisotropy due to the rippled topography. Earlier it has also been observed that c-axis of Co gets aligned along the long axis of nanowire or column due to coupling of magneto-crystalline and shape anisotropies that favors a coherent rotation. Co has an intrinsic magneto-crystalline easy axis along <002> direction. As revealed from RHEED measurement, this axis is textured, in the film plane along ripple direction. Therefore, both shape anisotropy and MCA gets coupled together in the same direction and enhances the strength of UMA.

To develop a clear understanding on the origin of this large UMA in the present sample, we annealed the sample from room temperature (RT) to 350ºC at an interval of 50ºC in the same UHV chamber. At each temperature the sample was annealed for 30 minutes whereas all the measurements were performed by cooling down the sample to RT. The variation of Hc at each annealing temperature is presented in Fig. 3(c). We observe that at temperatures around 250ºC, the difference between *Hc* value between two direction starts to decrease and disappears at around 350ºC. Furthermore, as shown in the inset of Fig. 3(c), the polar plot of Mr/Ms at this temperature has converted from dumble like shape into circular shape. The corresponding MOKE hysteresis loop in Fig 3(d) also displays identical hysteresis behavior. This confirms disappearance of anisotropy at around $350^0$C.

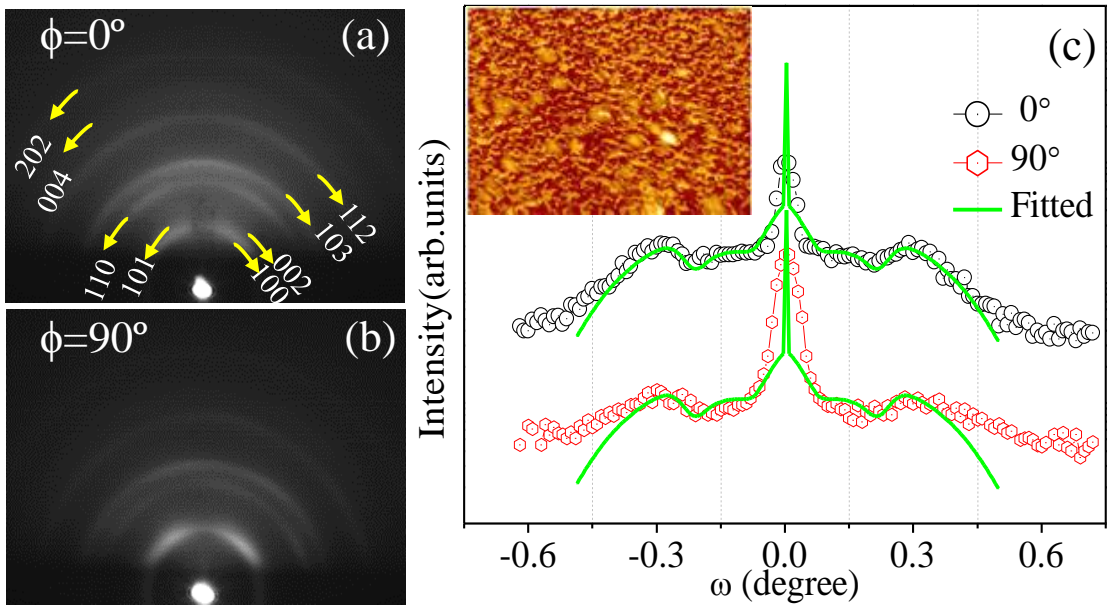

**Fig 4**: (a-b) RHEED images and (c) XDS curves taken along ϕ=0º and ϕ=90º after annealing for 30 minutes at 350ºC. Inset of (c) shows the corresponding AFM image.

The RHEED and XDS pattern taken at 350˚C along ϕ=0º and ϕ=90º is shown in Fig. 4(a,b) and 4(c) respectively. The RHEED pattern consists of symmetric Debye rings of uniform intensity distribution characteristics of isotropic polycrystalline nature of film. Thus, it is revealed that annealing removes the modulation of intensity of the rings due to annihilation of the texturing developed through this sequential etching and deposition method. In the same way the XDS measurement, as shown in Fig. 4c, exhibits symmetrically developed identical intensity pattern around the central peak. It confirms the relaxation of unidirectional nanopattern present on the surface due to migration of the Co atom. The corresponding surface topography of the film as imaged ex-situ by AFM is presented in the inset of Fig. of 4(c). No directional morphology is observed. From these results, we conclude that both the shape asymmetry and the preferential grain distribution is eliminated from the film by annealing. More importantly, it also confirms that the observed behavior of UMA is caused by both the film texture and morphology.

## Conclusions

Co film has been prepared by sequential deposition and IBE process and its morphology, surface texture and magnetic anisotropy has been characterized in-situ. The film grows in textured polycrystalline state and exhibits strong UMA, roughly one order larger than that similar kind of the previous studies. Detailed analysis reveals that the induced UMA not only originating from shape anisotropy but MCA due to crystallographic texture is also simultaneously contributing to enhance the strength of UMA. Therefore, the novelty of the present approach is the ability of imprinting film texture throughout the whole film layer along with surface morphology modification, consequently, not restricting the source of UMA limited to the thin surface region of a polycrystalline ferromagnetic film for flexible tuning and enhancement of the strength of UMA. These findings add a new route of inducing large UMA by IBE and suggest modeling studies with multiple components of magnetic anisotropy.